\documentclass[journal]{IEEEtran}

\usepackage{color}

\ifCLASSINFOpdf
  \usepackage[pdftex]{graphicx}

\else

\fi

\begin{document}

\title{10~Mb/s quantum key distribution}

\author{Z.~L.~Yuan, %~\IEEEmembership{Member,~OSA,}
        A.~Plews, R.~Takahashi, K.~Doi, W.~Tam, A.~W.~Sharpe, A.~R.~Dixon, E.~Lavelle, J.~F.~Dynes, A.~Murakami, M.~Kujiraoka, M.~Lucamarini, Y.~Tanizawa, H.~Sato, and A.~J.~Shields
\thanks{Z.~L.~Yuan, A.~Plews, W.~Tam, A.~W.~Sharpe, E.~Lavelle, J.~F.~Dynes, M.~Lucamarini, and A.~J.~Shields are with Toshiba Research Europe Limited, Cambridge Research Laboratory, Cambridge CB4 0GZ, United Kingdom email: zhiliang.yuan@crl.toshiba.co.uk}
\thanks{R.~Takahashi, K.~Doi, A.~R.~Dixon, A.~Murakami, M.~Kujiraoka, Y.~Tanizawa and H.~Sato are with Corporate Research and Development Center, Toshiba Corporation, 1 Komukai Toshiba-cho, Kawasaki-shi, Japan}% <-this % stops a space
\thanks{P\lowercase{ublished in} IEEE/OSA J. L\lowercase{ightwave} T\lowercase{echnol}. \textbf{36}, 3427(2018). Doi: 10.1109/JLT.2018.2843136.}
}

%\markboth{P\lowercase{ublished in} IEEE/OSA J. L\lowercase{ightwave} T\lowercase{echnol}. \textbf{36}, 3427(2018). Doi: 10.1109/JLT.2018.2843136.}%

\maketitle

\begin{abstract}
We report the first quantum key distribution (QKD) systems capable of delivering sustainable, real-time secure keys continuously at rates exceeding 10~Mb/s.
To achieve such rates, we developed high speed post-processing modules, achieving maximum data throughputs of 60~MC/s, 55~Mb/s, and 108~Mb/s for standalone operation of sifting, error correction and privacy amplification modules, respectively. The photonic layer of the QKD systems features high-speed single photon detectors based on
self-differencing InGaAs avalanche photodiodes, phase encoding using asymmetric Mach-Zehnder interferometer, and active stabilization of the interferometer phase and photon polarisation.
An efficient variant of the decoy-state BB84 protocol is implemented for security analysis, with a large dataset size of $10^8$ bits selected to mitigate finite-size effects. Over a 2~dB channel, a record secure key rate of 13.72~Mb/s has been achieved averaged over 4.4~days of operation.  We confirm the robustness and long-term stability on a second QKD system continuously running for 1~month without any user intervention.
\end{abstract}


\IEEEpeerreviewmaketitle

\section{Introduction}

Quantum key distribution (QKD) \cite{bb84,ekert91} provides a means for exchanging cryptographic keys, the security of which makes no assumptions about an adversary's computing power or technological capability.
Its potential for real-world use has stimulated a number of different implementations, ranging from prepare-measure to entanglement-based, weak-pulse to continuous variables, or free-space to fibre optics \cite{gisin02,scarani09,diamanti16}.
Constant efforts have been devoted towards extending the communication distance and the secure key rate (SKR): two key figures of merit defining the performance \cite{diamanti16}.
Recently, point-to-point QKD links have been deployed on installed fiber to form quantum networks \cite{qiu14}, which can be used to safeguard communication infrastructures that underpin today's society.

\begin{figure}
  \centering
  \includegraphics[width=0.5\textwidth]{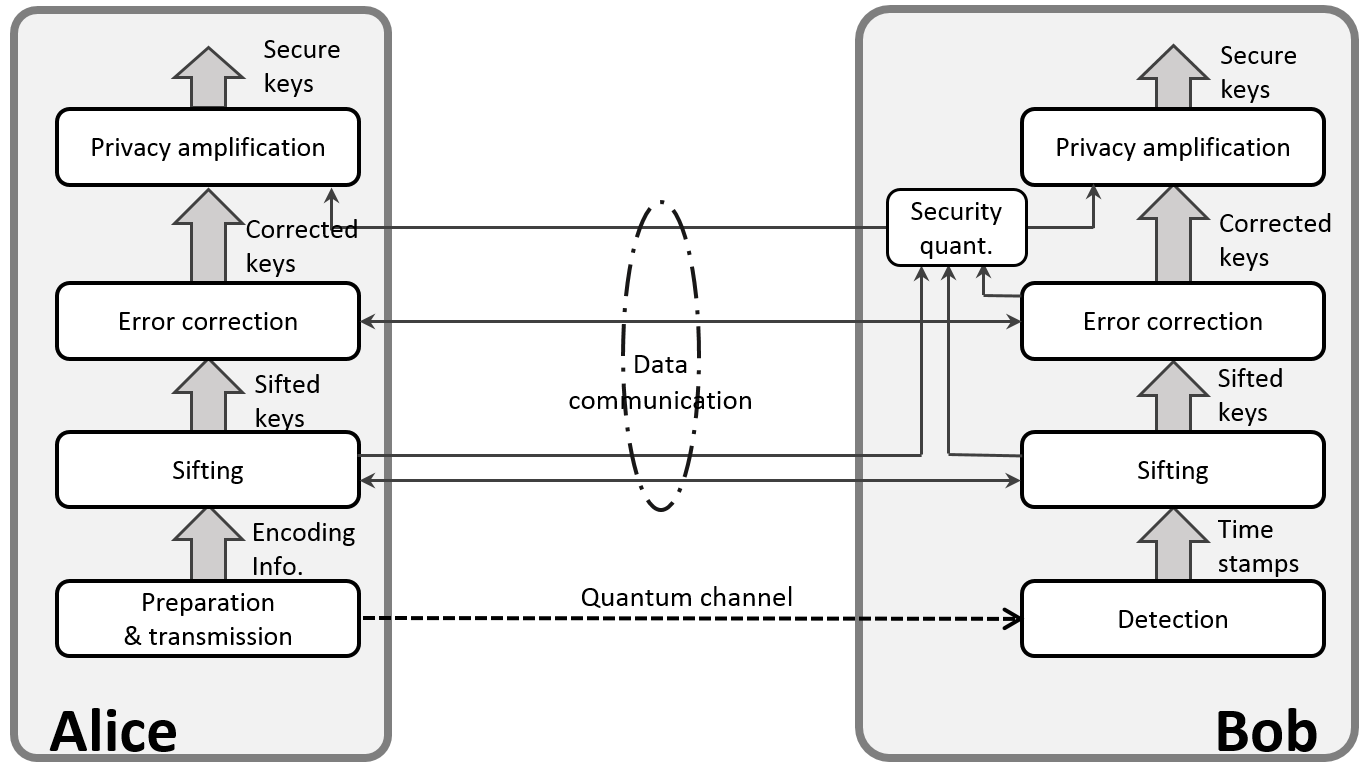}
  \caption{QKD data flow. }
  \label{fig:data_flow}
\end{figure}

Increasing the SKR is arguably the most pressing task in order to widen the applicable areas of the QKD technology.
Firstly, practical applications often require a minimum bandwidth, \textit{e.g.}, encryption of live speech requires a bit rate on the order of 10~kb/s while sharing human Genome data and distributed secure storage would desire a rate much higher than 1~Mb/s \cite{sasaki17}.
Secondly, a higher bandwidth would enable a network to provide simultaneous service to a larger number of users or user applications.
However, achieving a sustainable, high SKR requires improvements in all aspects of a QKD system, from protocol to hardware.
Figure~\ref{fig:data_flow} shows schematically the data flow among different layers in a QKD system.
It starts from the lowest, photonic layer dealing with preparation, transmission and detection of optical signals, followed by data post-processing layers including sifting, error correction (EC) and privacy amplification (PA), facilitated by an authenticated data communication channel.
Between EC and PA, there is a module for determining the amount of privacy amplification required in order to remove any information that might be known to an adversary (Eve), based on the security proof of the implemented QKD protocol.
As they are executed sequentially, the final secure key rate will be determined by the slowest layer.

The photonic layer has in the past limited the key rate.
With recent advances in single-photon detector technologies, the bottleneck has shifted upwards to the post-processing layers and the \textit{sustained} secure rate has improved steadily from several kb/s to a current record level of 1.9~Mb/s \cite{dynes16}.
A higher key rate has not been possible because of the speed limitation imposed by the data-processing layers.  We note that EC and PA were implemented in software running on general-purpose CPU's in the state-of-the-art system \cite{dynes16}.

Hardware acceleration has been recognized as the key to improve the post-processing throughput \cite{tanaka12,walenta14}. However, each layer requires different resources, adding to the complexity in hardware implementation.
Sifting requires high communication bandwidth as well as its efficient use in order to cope with exchanging reconciliation messages at sufficiently high rates, while its computational operation is simple and involves mainly memory look up and binary bit comparison.
In contrast, EC and PA are computationally intensive, with the latter requiring significant memory resources to process datasets of very large size ($10^8$ bits) in order to mitigate statistical finite size effects.

Here, we report the first QKD systems capable of delivering sustainable secure keys at a rate in excess of 10~Mb/s.   This achievement was enabled by our development of sifting and control electronics, field-programmable gate array (FPGA) based low-density parity check (LPDC) EC and co-processor based PA modules.  Together with a high speed photonic layer, this new hardware has enabled a sustainable, real-time SKR of $13.72\pm0.74$ Mb/s over a channel loss of 2~dB, equivalent to 10~km fiber. The system robustness has been confirmed by one month continuous operation of a second system over a 2~dB quantum channel formed by fiber and delivering an average SKR of 11.53~Mb/s.

\section {Requirements for high secure key rates}

In a qubit-based QKD system, the obtainable SKR ($R$) between the transmitter (Alice) and receiver (Bob) can be expressed as
\begin{equation}\label{eq:SKR}
R = f \eta_d \eta_{sift} \left \{\underline{p_1} \left [1 - h \left (\overline{e_1} \right ) \right ] - f_{ec} h(e) \right \},
\end{equation}
\noindent where $f$ is the system clock frequency, $\eta_d$ is the detection probability of signal pulses, $\eta_{sift} < 1 $ the sifting coefficient which is defined as the proportion of the system time slots that contribute to generation of sifted keys,
$\underline {p_1}$ is the lower bound for the probability that a sifted count arises from the detection of a single photon state prepared by Alice,
$\overline{e_1}$ is the upper bound of the phase error rate of the single photon states, and $e$ is the quantum bit error rate (QBER) in the sifted key. $h(x) = - x\log_2 x - (1-x) \log_2(1-x)$ is the binary Shannon entropy, representing the fraction of information that has to be revealed by a perfect EC algorithm in order to correct a binary data block with an error ratio of $x$.  $f_{ec}$ ($>1$) is the co-efficiency of the actual employed EC algorithm, and $f_{ec} h(e)$ represents the amount of information disclosed by it.

Inferring from Eq.~\ref{eq:SKR}, we may list below the requirements for high SKRs:
\begin{enumerate}
  \item A photonic layer of transmitting and detecting optical pulses at high clock rates ($f$) and high detection probability ($\eta_d$);  $\eta_d$ here includes the quantum channel loss.
      For fiber-optical QKD systems, their SKRs decrease exponentially with the communication distance. Hence, high SKRs favour short or low-loss fiber channels.
  \item The sifting electronics that can handle a raw count rate greater than $f\eta_d$;
  \item A QKD protocol that enables efficient extraction ($\eta_{sift}$) of sifted keys from raw photon detection events;  We note that the standard BB84 protocol features $\eta_{sift} \leq 0.5$.  An efficient protocol allows also tight bounds for parameters $\underline{p_1}$ and $\overline{e_1}$.
  \item A robust control hardware that is able to maintain the alignment between the transmitter and receivers' optical apparatus so as to have a low QBER ($e$) and a high proportion of time slots for transmitting quantum keys;
      time slots used for stabilising the system should be kept minimal.
  \item An EC implementation that has a low information leakage ($f_{EC}\rightarrow 1$);
  \item A PA implementation that must be able to handle large dataset size to mitigate finite-size effects arising from statistical fluctuations in the measured quantities;
  \item Finally, both EC and PA modules must have data throughput greater than the sifted key rate ($f \eta_d \eta_{sift}$).
\end{enumerate}

\noindent To achieve 10~Mb/s secure key rates would require EC/PA throughput of 40~Mb/s and sifting throughput of 50~MC/s, when considering typical experimental parameters found in previous QKD implementations \cite{dynes16}.

\section {Protocol}

Early QKD security proofs were developed in the so-called ``asymptotic scenario'', where it is assumed that an infinite dataset is available to the experimenters, who can thus determine the QKD parameters with infinite precision.
This is clearly unphysical.
In a real situation, the dataset is always finite and the measurement precision is therefore limited by the statistical fluctuations in the sample.
To correct this, we introduced in 2012 the ``T12'' protocol \cite{lucamarini13}, which features composable security against collective attacks in the finite-size scenario and provides high key distribution rates.

In the T12 protocol, the transmitter (Alice) prepares at random one out of four quantum states indicated as $|0\rangle_Z$, $|1\rangle_Z$ (\textit{majority} and \textit{data} basis \textit{Z}) and $|0\rangle_X$, $|1\rangle_X$ (\textit{minority} and \textit{test} basis \textit{X}), similarly to the well-known BB84 protocol.
The bases are selected with probability $p_Z \geq 1/2$ and $p_X=1-p_Z$ and the secure key rate of the protocol is given by the sum of the rates distilled separately in the two bases. This makes it possible to optimize the parameter $p_X$ to achieve the highest possible key rate. An optimal choice is often found to be $p_X < 1/16$ \cite{lucamarini13}, which entails that only a small fraction ($2p_Xp_Z < 0.125$) of detected counts are discarded due to non-matching bases.

For each state, one of three photon fluxes, \textit{u} (signal), \textit{v} (decoy) or \textit{w} (vacuum), are randomly selected with probabilities $p_u$, $p_v$ and $p_w$, respectively, to enable the implementation of the decoy-state technique \cite{lo05,wang05}.
The decoy-state estimation routine is carried out numerically for every dataset acquired by the system, making it possible to optimise the values of the photon fluxes and their preparation probabilities.

The T12 protocol can easily tolerate security parameters as small as $10^{-10}$, even in presence of dataset sizes as small as $10^5$, although much larger dataset sizes are necessary in order to reduce statistical finite-size effects and allow efficient key generation. We choose our dataset size as 100~Mb for our QKD systems.  This choice allows 85\% of the asymptotic secure key rate, and remains manageable by our privacy amplification module which will be described later.
For a system generating 100-Mb keys every 100 seconds, the security parameter ($10^{-10}$) corresponds to a failure probability equal to one single event in 30,000 years. Despite such a strict security parameter, the T12 protocol provides a typical secure key rate in excess of 1~Mb/s over an optical fibre length of 50 km \cite{lucamarini13}.
Record rates were reported using this protocol for several distances up to 240~km \cite{frohlich16}, even in presence of the noise introduced by classical channels multiplexed with the quantum channel~\cite{dynes16}.

The security in the protocol can be rigorously quantified by relating it to the size of the experimental data sample. The simplicity of the protocol further allows to incorporate additional assumptions of the QKD theory which are not currently met in the implementation, thus reducing the existing gap between these two aspects of QKD \cite{lucamarini15,dynes18}. For the same reason, it can be easily exported to other systems and situations in which a worst-case analysis becomes compelling.

In the present QKD systems, the T12 protocol is implemented with operation parameters listed in Table~\ref{tab:t12}, with aim to achieve high SKRs at short fiber distances.  Here, $Z-$basis counts are used for key distillation while the full information of $X$-basis counts is revealed for phase error estimation.   The sifting efficiency is 0.90, obtained from

\begin{equation}\label{eq:sifting_eff}
\eta_{sift} = (1-p_{st}) \cdot p_u \cdot p_Z^2,
\end{equation}

\noindent where $p_{st}$ is the fraction of time slots that are used for phase stabilization.
Such sifting efficiency is almost twice as efficient as the standard BB84 protocol.  With a typical QBER of 3\% and PA dataset size of $10^8$ bits, its PA compression ratio is found to be about 0.29, \textit {i.e.}, on average 0.29 secure bits can be distilled per bit in the error-corrected key.

\begin{table}[t]
\caption{List of parameter settings in the implemented T12 protocol. $f$: system clock frequency; $u,v,w$ and $p_u, p_v, p_w$: photon fluxes and respective probabilities for signal, decoy and vacuum pulses; $p_Z, p_X$: $Z$ and $X$ basis probabilities; $p_{st}$: probability of pulses used for active phase compensation; $\epsilon$: protocol security parameter. }
%\begin{ruledtabular}
\begin{tabular}{c|c}
  \hline\hline
  Parameter & Setting \\
  \hline\hline
  $f$  & 1~GHz \\
  \hline
  $u$, $v$, $w$  &  0.4, 0.1, 0.0007 \\
  \hline
  $p_u, p_v, p_w$  & 96.973\%, 1.661\% 1.466\% \\
  \hline
  $p_Z$, $p_X$  & 96.677\%, 3.323\% \\
  \hline
  $p_{st}$ & 1/128 \\
  \hline
  PA dataset & 100.66~Mb \\
  \hline
  $\epsilon$ & $10^{-10}$ \\
  \hline\hline
\end{tabular}
%\end{ruledtabular}
\label{tab:t12}
\end{table}

\section {System hardware}

\begin{figure}[b]
  \centering
  \includegraphics[width=0.5\textwidth]{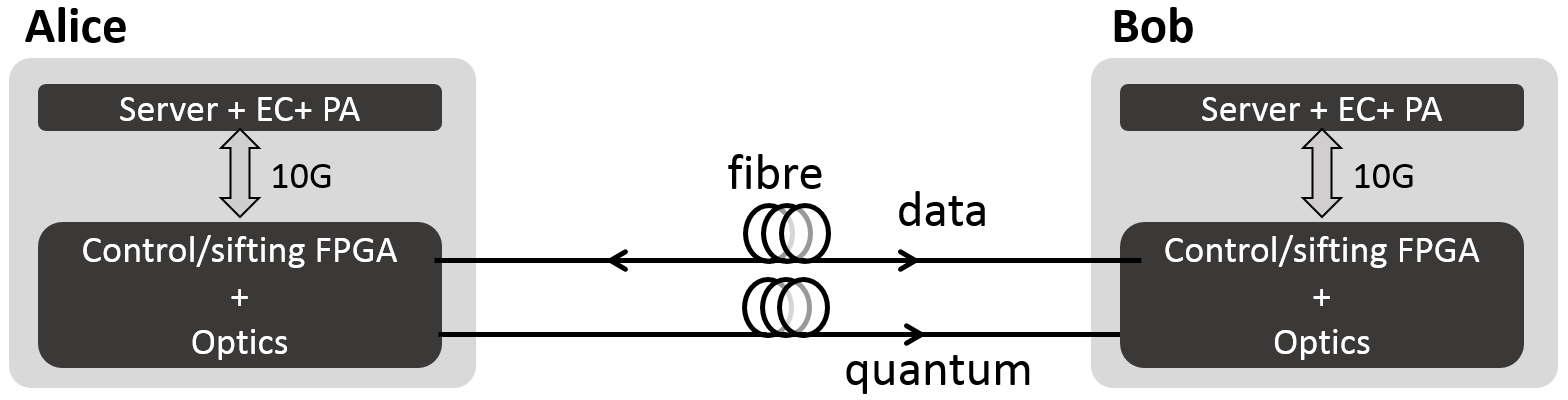}
  \caption{Schematics for experimental setup. Quantum transmitter/receiver is housed in a 2U 19-inch rack unit. 10G communication interfaces are chosen to handle and route all classical communications. The EC/PA hardware accelerators are housed inside the control servers.}
  \label{fig:setup}
\end{figure}

Figure~\ref{fig:setup} shows the schematics of the QKD hardware. Quantum transmitter and receiver are housed in 19-inch 2U rack units, consisting both optics and control/sifting electronics, while EC/PA hardware are hosted in the 1U servers.
Standard 10G Ethernet interfaces are used for local communication between server and sifting electronics or for remote communication between Alice and Bob's sifting electronics.

\subsection {Optics}
\begin{figure}
  \centering
  \includegraphics[width=0.5\textwidth]{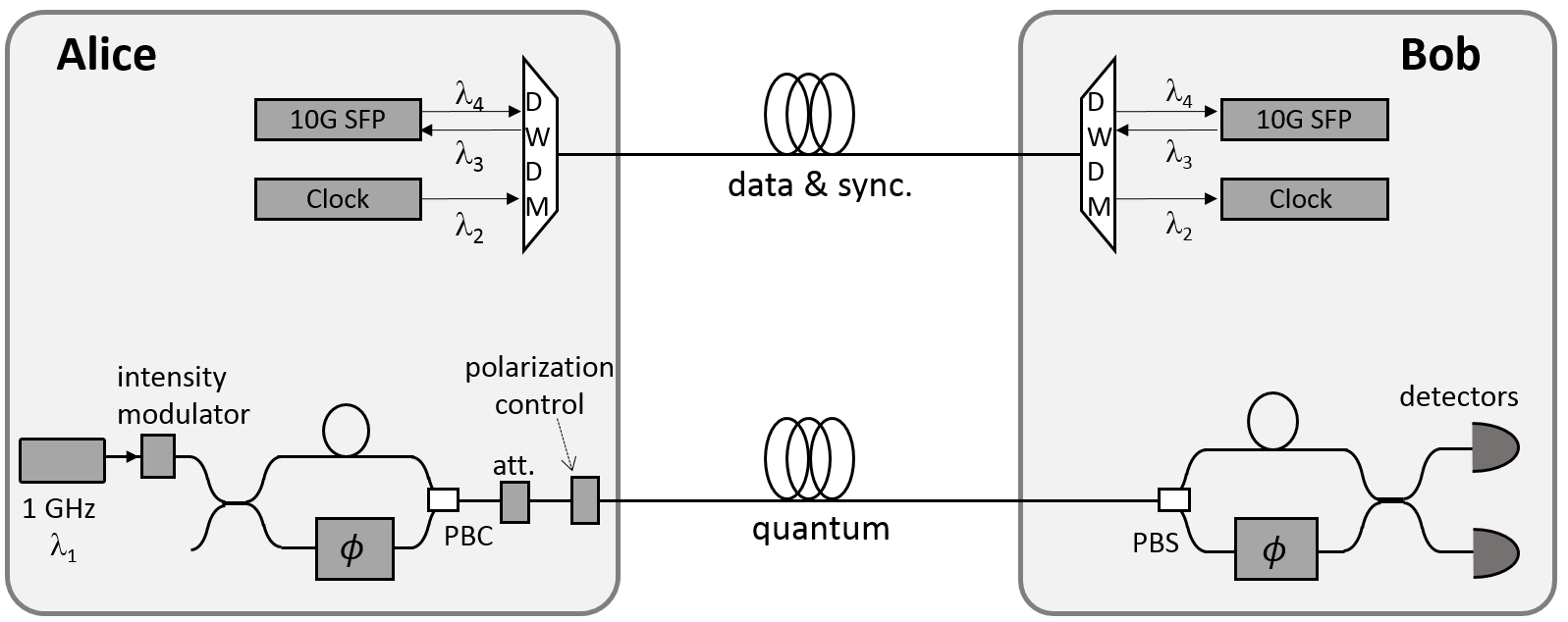}
  \caption{Optics.  Dark shaded components are active optical components that are controlled by sifting/control electronics.    DWDM: dense wavelength division multiplexing module; PBC/PBS: polarisation beam combiner/splitter; Att: electrically controlled variable optical attenuator.}
  \label{fig:optics}
\end{figure}

Figure~\ref{fig:optics} shows the optical layout of the QKD system.
Four optical wavelengths ($\lambda_1$ -- $\lambda_4$) are used to transmitting optical signals between the QKD transmitter (Alice) and receiver (Bob).  Three wavelengths are assigned to classical signals and transmitted through a single fiber using standard DWDM 100G-spaced multiplexers.   The classical channels include a 10~Gb/s bidirectional data link for QKD sifting and reconciliation ($\lambda_3$: 1529.55~nm, $\lambda_4$: 1528.77~nm) and a unidirectional, pulsed transmitter-receiver pair ($\lambda_2$: 1531.22~nm) for clock synchronisation. The quantum signal ($\lambda_1$: 1550.12~nm) is presently transmitted through a separate fiber, but can be integrated into the same fibre that carries data in future.

The quantum sub-system implements the T12 protocol using optical pulses of three different intensities.
A gain-switched DFB laser diode, in combination with an intensity modulator and optical attenuator, is used to generate these optical pulses at the wavelength of  1550.12~nm ($\lambda_1$) and a clock rate of 1~GHz, with their fluxes set at three different values to implement the decoy state technique, 0.4 (stabilization and signal), 0.08 (decoy) and 0.0007 (vacuum) photons/pulse. Gain-switching of the laser ensures global phase randomization \cite{yuan14,kobayashi14} which is a precondition to apply decoy-state analysis, while use of a single laser source prevents encoding side-channels.
Quantum information is encoded upon optical phase using an asymmetric Mach-Zehnder interferometer (AMZI), which splits each input pulse into a pair of pulses of orthogonal polarizations.

At the receiver's side, a matching AMZI decodes the phase information and directs the quantum signals into two room-temperature InGaAs avalanche photodiodes operating in self-differencing mode \cite{yuan07,comandar14}. The detectors are temporally aligned so that only the photons that have passed through the short arm of one AMZI and the long arm of the other AMZI are detected. These detectors feature ultrashort detection deadtime and $>100$~MC/s count rates. The receiver's interferometer loss is about 2~dB.

In order for continuous operation, the quantum sub-system needs to maintain the photon polarisation, the interferometer phase delay, and the photon arrival time  because the effects of environmental changes in both the transmission fibers and the QKD unit locations
can cause disturbances in the transmitted quantum states.
The active stabilization is realized through applying compensation signals to various active components.
Alice applies the polarisation control using her electrically driven polarization controller, while Bob performs phase stabilization and detector gate delay optimization.

\subsection {Control and sifting electronics}

Custom-made FPGA boards were developed to control the QKD optics and handle the signal sifting.  Alice and Bob's boards are identical in terms of physical hardware, but differ in their loaded programmes. They handle optical signal modulation, photon detection time tagging, active stabilisation of optics, random number generation, sifting and packet data communication/routing.  Each FPGA board has a 10G SFP+ interface (850~nm) connecting locally to its control server, and a second 10G SFP+ interface ($\lambda_2$, $\lambda_3$) for remote communication via the data fiber link with its peer FPGA board. As can be seen in Fig.~\ref{fig:setup}, these 10G links form a daisy chain to carry all data communications between different modules of the QKD system.

Alice's FPGA board provides high speed driving signals at a clock rate of 1~GHz to the quantum laser, phase modulators and intensity modulator. Digital delay lines are used to temporarily align these signals in order to achieve desired modulation to the quantum signal. Alice's FPGA board transmits clock signal optically to Bob.  Bob's board produces modulation signals similarly to Alice's, but additionally implements high speed time-tagging to record incoming photon detection events. To meet the requirement by T12 protocol, the basis and decoy modulation channels are designed to have selectable settings for the probabilities of $p_X$, $p_u$, $p_v$ and $p_w$.

We use active stabilization to compensate any drift in the ambient conditions. The photon polarisation and arrival time are corrected by maximizing the total detection rates through applying DC voltages to the electrically-driven polarisation controller at Alice and the gate delays to single photon detectors at Bob, respectively.   To stabilize the interferometer phase, the count rate of the unmodulated stabilization pulses is used as a feedback to adjust the voltage that is applied to the DC input to Bob's phase modulator.

Sifting, including both basis reconciliation and decoy-state statistics collection, is completed in the hardware level with a proved capacity of 60~MC/s.

\subsection {Error Correction module}

Low Density Parity Check (LDPC) based EC \cite{gallager62, mackay99} was selected to satisfy the requirement for high throughput ($>$40~Mb/s) post processing.
LDPC codes have several advantages in this regard. They have very low communication complexity, requiring only a single unidirectional message between the transmitter and receiver, with the result that communication latency does not reduce the throughput. This is in contrast to protocols such as Cascade \cite{brassard94} which require many round trip messages.

LDPC decoding algorithms are also widely studied and implemented, with optimised algorithms offering good compromises between computational complexity, revealing extra redundant information and failure probability \cite{mackay99,fossorier99}. For high throughput algorithms with lower computational complexity are generally used, with some trade off in the extra information revealed and so final secure key rate.

The structure of the decoding algorithm, combined with the simple communication complexity, also allows for the EC process to be readily parallelised. This is an important factor for achieving high decoding throughput, and also makes LDPC decoders suitable for implementing using graphical processing units (GPU) and FPGA for very high throughput \cite{dixon14,walenta14}.
We have previously realized a GPU implementation, whose throughput is slightly below the requirement for a 10~Mb/s QKD system \cite{dixon14}.
Therefore, we choose here to implement the LDPC algorithm onto a pair of FPGA cards, each of which is plugged into a PCIe slot in Alice's or Bob's server.  Alice's card parallelizes syndrome generation while Bob's card handles decoding.

\begin{figure}
  \centering
  \includegraphics[width=0.42\textwidth]{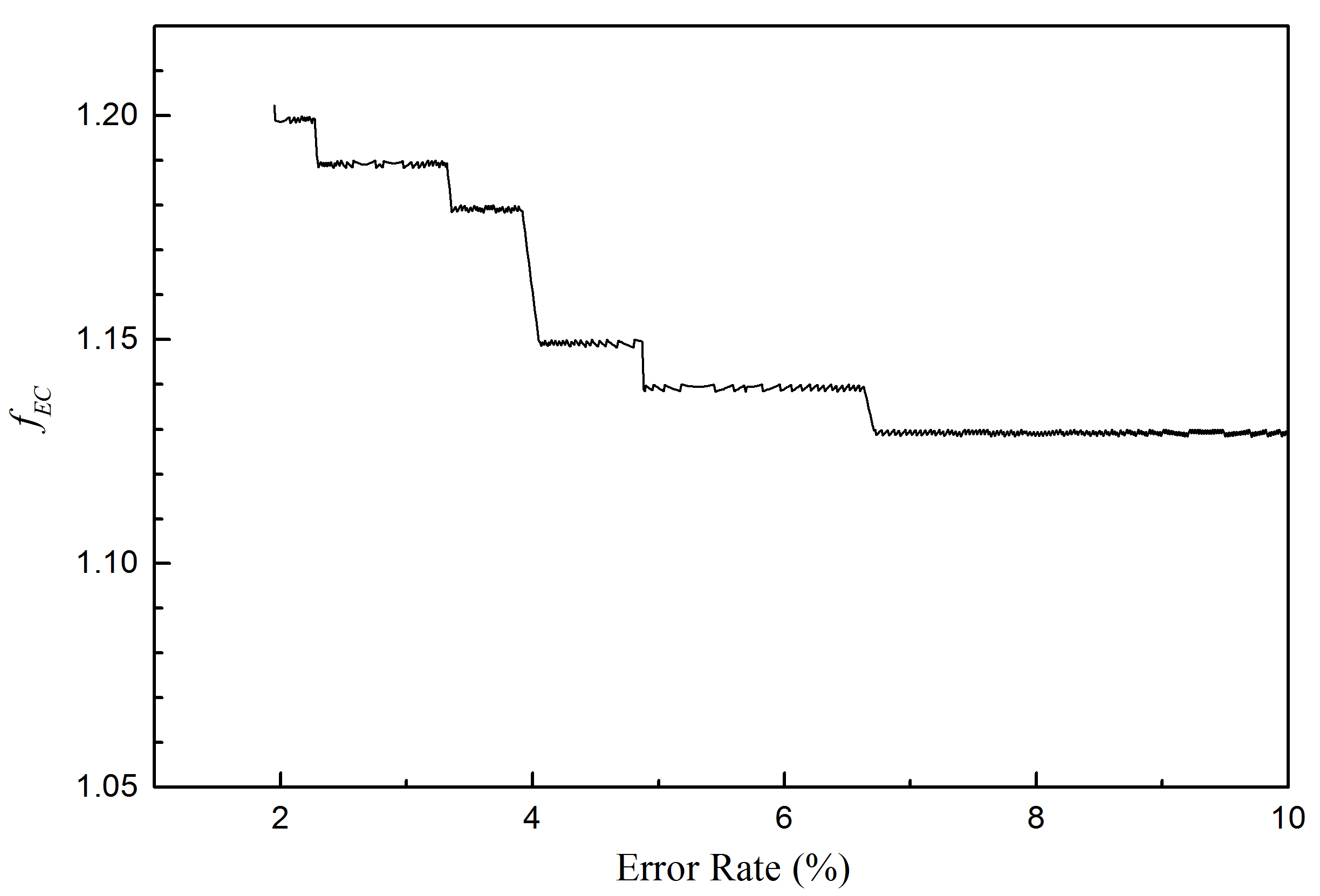}
  \caption{EC efficiency as a function of QBER for the LDPC-EC implementation based on FPGA. }
  \label{fig:ldpc1}
\end{figure}

We select quasi-cyclic LDPC code for its compact implementation, and proposed a new decoding routine \cite{doi15} in order to prevent performance deterioration in its FPGA implementation. We implement code rate adaptive modulation so as to achieve an optimal EC efficiency.
The code rate, and hence the amount of information disclosed by EC, is determined by the QBER in the EC data block.   We determine the EC efficiency of the FPGA-LDPC implementation to be from 1.13 to 1.20 for QBER's between 2\% and 10\%, as shown in Fig.~\ref{fig:ldpc1}.

The above efficiencies are achievable only when the number of errors in the sifted data block are precisely known.
Processing real QKD data, the efficiencies will become slightly poorer as will be reported later in Section IV,  because the QBER is not precisely known before EC taking place. The LDPC code rate and the syndrome data calculation have to rely on an estimated QBER, obtained via revealing a small sample in the sifted data block.
An underestimation of QBER will increase LDPC decoding failure probability, and to avoid this a margin on top of the estimated QBER is used to determine the LDPC code rate and therefore reduces the EC efficiency.

In the FPGA-LDPC implementation, we set each EC data block to be 1~Mb, of which 8 kb is revealed for error estimation.   The remaining data is divided into sub-blocks to be error corrected in parallel. If decoding of all sub-blocks is verified to be successful, the corrected data block is transferred to the PA process. Otherwise, the whole block of the sifted key is discarded. Our test reveals the EC throughput to be 55~Mb/s, sufficient to support 10~Mb/s SKRs.

\subsection {Privacy amplification module}

Privacy amplification (PA) is an essential post-processing step in QKD. Its function is to compress a length of error-corrected key into a shorter one thus removing information known to Eve via her interception of  either optical transmission, error correction stages or both.
The simplest approach is direct matrix multiplication, which has computational complexity of $O(n^2)$.
It is feasible with CPU implementations for small dataset sizes ($\leq 10^5$ bits), but rapidly becomes troublesome with larger dataset sizes.
FPGA implementations have often been used to speed up PA process with a dataset size of up to $10^6$ bits \cite{tanaka12, zhang12, walenta14, yang17}. However, such dataset size is still two orders of magnitude smaller than $10^8$ bits, a desirable size \cite{lucamarini13} to mitigate statistical finite-size effects.

To overcome this problem, we first apply number theoretical transform (NTT) technique \cite{agarwal75} to Toeplitz matrix multiplication, a transformation that belongs to the family of \textit{universal-2} hash functions.
This technique reduces the computational complexity from $O(n^2)$ to $O(n\log n)$.  We note that an alternative PA approach \cite{zhang14}, \textit{i.e.}, Sch\"onhage-Strassen algorithm based multiplication of large integers, has a good but slightly higher computational complexity of $O(n \log n \log \log n)$.
A CPU implementation of the NTT algorithm has given a PA throughput of 28.22~Mb/s at a dataset size of $10^8$ bits \cite{takahashi16}. While
being a significant improvement over direct matrix multiplication, it remains insufficient for 10~Mb/s secure key rates.

We then resort to exploiting massive parallelism offered by Intel\textsuperscript{\textregistered} Xeon Phi\textsuperscript{TM} coprocessors to further improve the PA throughput.
In order to efficiently utilize the coprocessor for matrix multiplication, we apply several programming techniques: vectorization for matrix transpose and butterfly computation of NTT, suitable instruction set regarding cache hit ratio, loop unrolling to butterfly computation for reducing the number of iterations, parallelization by multi-thread processing of data input and output, and parallelization of matrix multiplication and secure key length estimation.

The implementation supports a variable dataset size up to $2^{27}$~bits (134~Mb), thanks to the reduced computational complexity by NTT implementation.
We note that $\sim$100~Mb dataset size is necessary to give a high secure key rate owing to finite-size effects in the extraction of secure keys. At this dataset size, our evaluation on the NTT implementation gives a throughput of 108.77~Mb/s. This throughput can support a secure key rate exceeding 20~Mb/s in the T12 protocol even when the QBER is 5\%.

In our QKD systems, we choose  100.66~Mb ($96 \times 1024 \times 1024$ bits) as the PA dataset size and the PA module allows a maximum compression ratio of 1/3.

\section{System evaluation and performance}

\begin{figure}
  \centering
  \includegraphics[width=0.5\textwidth]{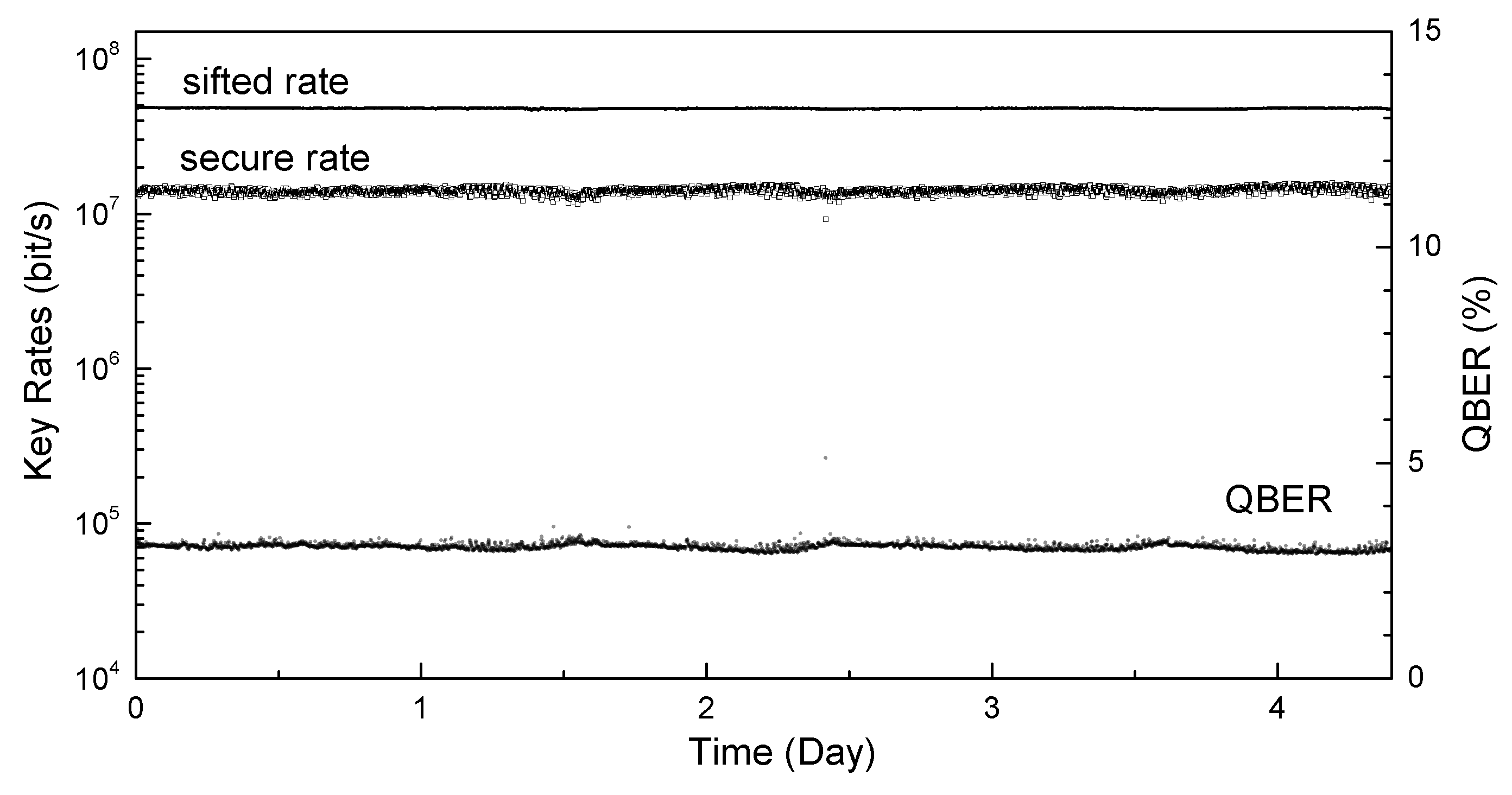}
  \caption{Sifted and secure bit rates and the QBER as a function of time, obtained from 4.4-days continuous operation of System-I. The quantum channel is formed by a short spool of fibre and its channel loss is set to be 2~dB, equivalent to the loss of 10~km standard single mode fiber with a loss rate of 0.2~dB/km.}
  \label{fig:keyrate}
\end{figure}

We constructed two QKD systems which were subsequently used for evaluating their secure key rates. In their evaluations, the quantum channel loss was set to 2~dB, equivalent to the loss of 10~km standard single mode fiber with a loss rate of 0.2~dB/km.
All sifting, EC and PA processes are pipelined and performed in real time with secure keys written to hard drives.
The QKD control software was configured to log control, sifting, EC and PA parameters for off-line analysis.

Systems I and II ran continuously for 4.4 days and 1 month, respectively, without any user intervention during the evaluations.

\subsection{Record secure key rate}

System I is equipped with two room-temperature InGaAs single photon detectors featuring a detection efficiency of 31\%, a combined dark count rate of 450~kHz and an afterpulsing probability of 4.4\%.
In the test, its 2~dB quantum channel was formed by a 400~m fiber spool and additional attenuation from an optical attenuator.

Figure~\ref{fig:keyrate} shows the sifted and secure rates and QBER.
All these data were reported at an interval of every 100.66~Mb successfully reconciled keys or approximately 2~seconds duration, obtained from the PA routine.
The system reports stable sifted key rate, measured to be $47.83 \pm 0.22$~Mb/s with a standard deviation of 0.5\%, and a stable QBER of $(3.07\pm0.05$)\%, as shown in Fig.~\ref{fig:keyrate}.  The detector afterpulsing makes the biggest contribution (2.2\%) to the measured QBER.

\begin{figure}
  \centering
  \includegraphics[width=0.42\textwidth]{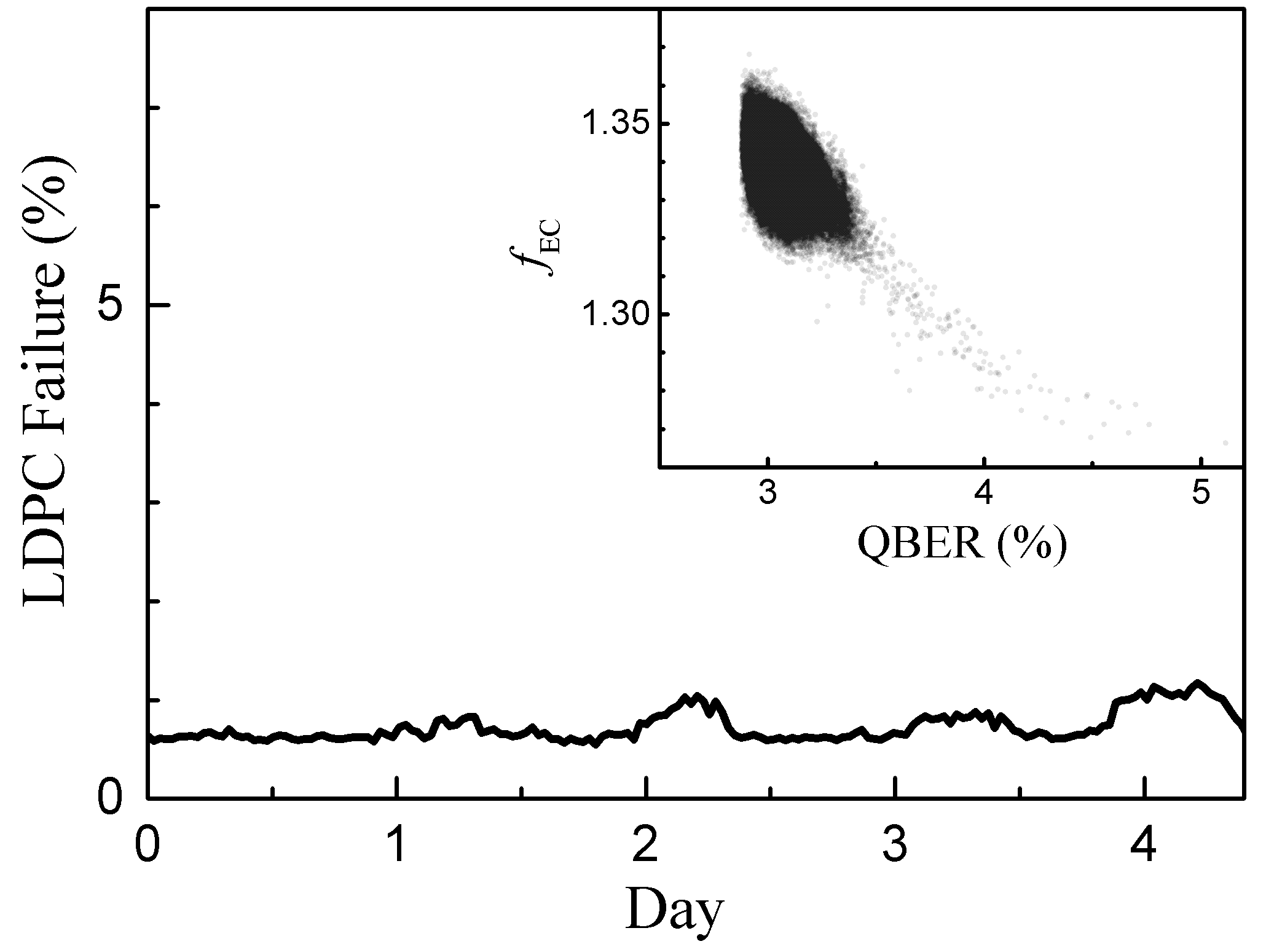}
  \caption {Low-density parity-check error correction (LDPC-EC) failure probability. The inset shows the EC information leakage, compared with the theoretical minimum.}
  \label{fig:ldpc}
\end{figure}

LDPC-based EC suffers from a non-zero decoding failure probability, and the EC control software performs a verification process after correcting each EC data block. Once an EC failure is confirmed by its verification procedure, the entire data block is discarded and does not enter the subsequent PA process.
We determine the failure probability to be ($0.73\pm0.15$)\%, as shown in Fig.~\ref{fig:ldpc}.
The inset shows the EC coefficient as a function of QBER.  Unlike the theoretical result in Fig.~\ref{fig:ldpc1}, $f_{EC}$ is not a singular value for each QBER value.   This is because the LDPC code rate and hence the corresponding $f_{EC}$ value are determined by the estimated QBER.
At 3.0\% QBER, $f_{EC} = 1.32 - 1.36$, which is slightly less efficient than the value of 1.19, obtained in Fig.~4 with a precisely known QBER.
This degradation is attributed to the inevitable error in the estimation of QBER.

\begin{figure}[b]
  \centering
  \includegraphics[width=0.42\textwidth]{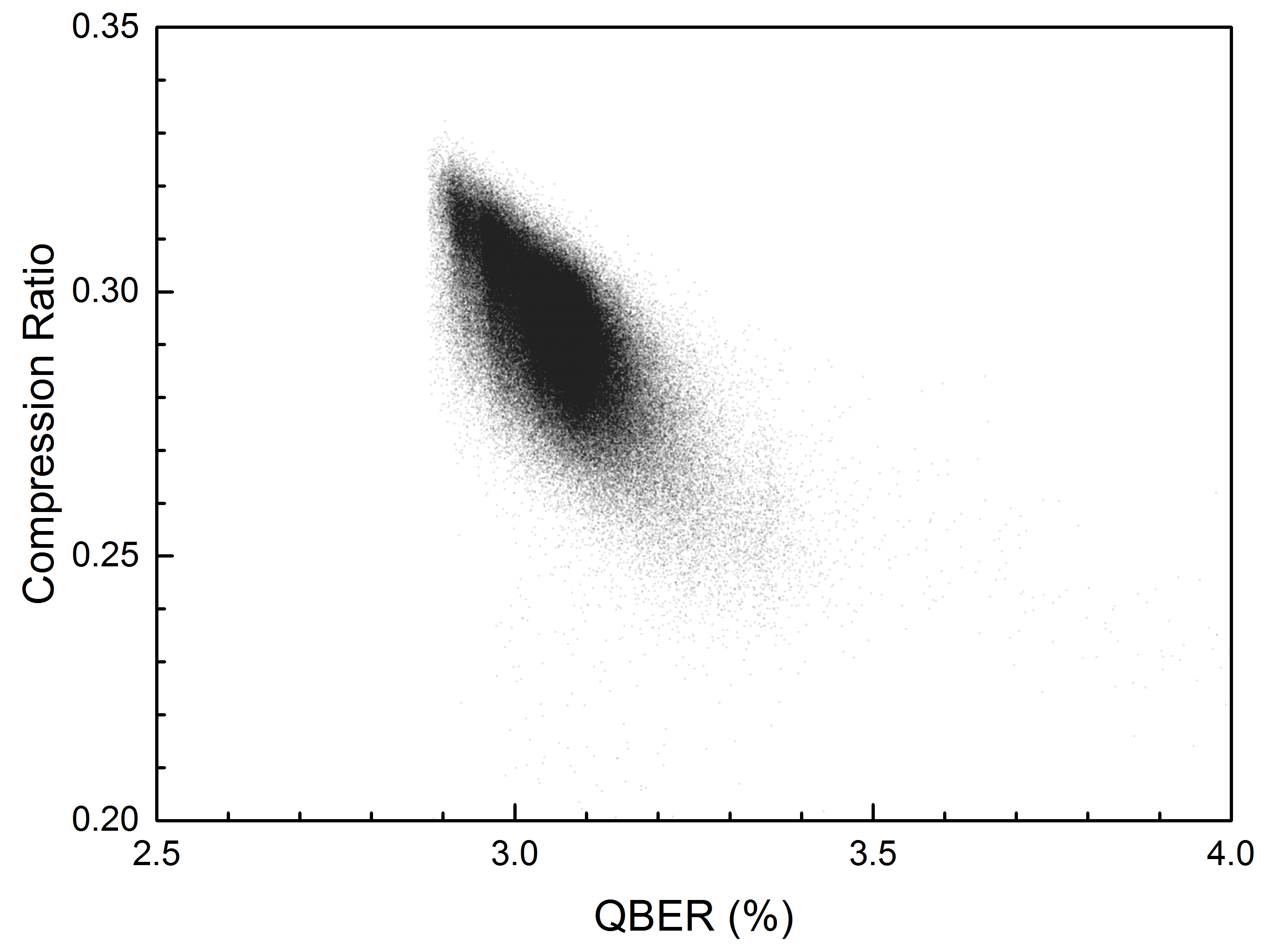}
  \caption {The PA compression ratio \textit{vs.} QBER obtained from the 4.4 days test.}
  \label{fig:pa_ratio}
\end{figure}

Despite the degraded EC efficiencies,  it is still possible to achieve a high PA compression ratio, \textit{i.e.}, the ratio of the number of secret bits to the number of error corrected bits, thanks to the low QBER.  Figure~\ref{fig:pa_ratio} shows the compression ratio obtained during the test period, measured as $0.292 \pm 0.016$.  On average, the system distills 29.39~Mb secure bits per 100.66~Mb error-corrected keys. We show the secure key rate a function of time in Fig.~\ref{fig:keyrate}.
Its value fluctuates slightly stronger than the sifted key rate, because of its sensitivity to the QBER and the decoy counts statistics.  Nevertheless, all the secure rate values, except for a few data points, stay well above 10~Mb/s.   The fluctuation in the secure key rate is measured to be 5.4\%.
From the actual amount of secure key materials written into the hard drives, we determine the average secure key rate to be 13.72~Mb/s for the test duration, with a total of 5.2~Tb key materials generated. This is the first time for any QKD system to ever obtain a secure key rate exceeding 10~Mb/s.

\subsection {Long-term system stability}

System II is identical to the first one, but differs in the performance of single photon detectors. Its detectors have a slightly lower  detection efficiency (28\%) than those in the first system, but have comparable afterpulsing and dark count rate performance.

\begin{figure}
  \centering
  \includegraphics[width=0.5\textwidth]{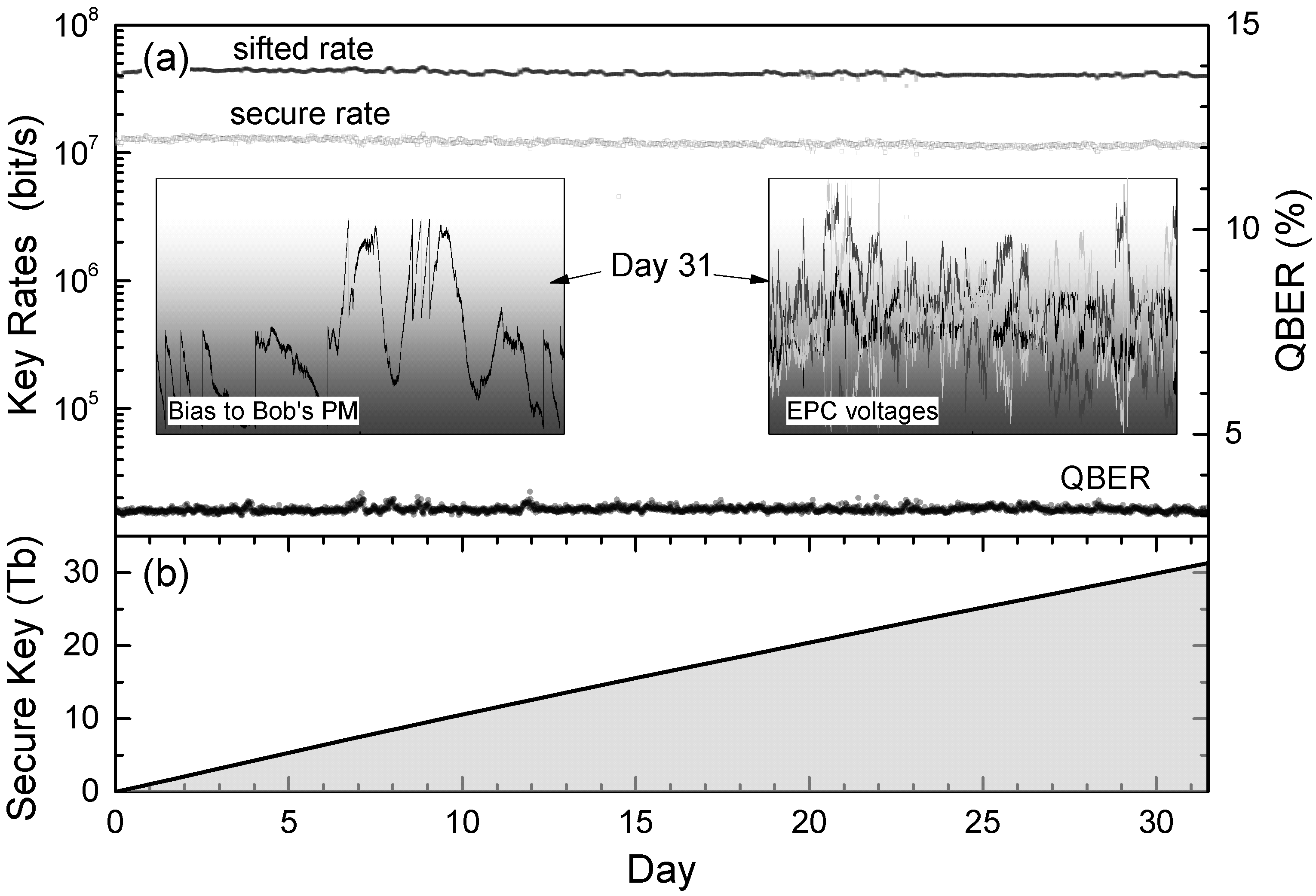}
  \caption{One month continuous operation of System II. (a) Sifted and secure key rates and the QBER;  Insets: 24-h records of the DC bias values applied to Bob's phase modulator (left) and Alice's electrically driven polarization controller (right).  (b) Total amount of secure key bits accumulated.}
  \label{fig:1mth}
\end{figure}

For better evaluation of the system stability,  it is desirable to use long fiber spools
because they provide a more realistic scenario of photon polarisation drift and fluctuation.
As we did not have 10~km fiber spools available at the time, two 25~km fiber spools were used instead, one for the quantum channel and one for data communication and synchronization. Each 25~km fiber spool has a loss of 5~dB.
To maintain the same quantum channel loss as System I, we compensate this extra loss by increasing the photon fluxes leaving Alice's unit. It is equivalent to view Alice's system consisting of 15~km of the total fiber spool, and the quantum channel is formed effectively by the remaining 10~km fiber with 2~dB loss. We stress that this approach does not weaken our result as compared with using 10~km fiber spools, because extra fiber can only deteriorate the system performance due to the increased polarisation instability.

We subject System II to a 1 month test for evaluating the robustness of the system, with the result shown in Fig.~\ref{fig:1mth}.  The sifted and secure key rates and the QBER are measured to be $42.21 \pm 1.65$~Mb/s, $11.53 \pm 0.65$~Mb/s, and $3.16 \pm 0.07$\%, respectively.
The fluctuations of the measured quantities are slightly larger than System I, due to the extra instability from the 25~km fiber spool.  However, all these quantities show excellent stability, illustrating the robustness of the active stabilisation routine.
The insets of Fig.~\ref{fig:1mth} show the voltage biases applied to Bob's phase modulator and Alice's electrically driven polarisation controller over 24 h on the 31$^{st}$ day.  The key rates are about 16\% lower than System I, which is attributed to the lower detection efficiencies of the InGaAs detectors and a slightly higher QBER.  Nevertheless, the secure bit rate has stayed over 10~Mb/s over the entire period.

A total of 31.35~Tb secure key materials were generated in this test, see Fig.~\ref{fig:1mth}(b).

\section{Conclusion}

We have developed and successfully demonstrated high bit rate QKD systems capable of generation of real-time secure keys at a sustainable rate exceeding 10~Mb/s.  To achieve this rate,  we have made remarkable improvement on the data throughput by successful development of QKD post-processing modules, with respective individual maximum throughput of 60~MC/s, 55~Mb/s and 108~Mb/s for sifting, error correction and privacy amplification.   We have integrated these modules into a compact QKD system, and achieved a record secure key rate of 13.72~Mb/s over 2~dB loss channel, equivalent to 10~km standard telecom fiber.   The system robustness on real fiber is confirmed with a 1-month continuous operation over a quantum channel with its 2~dB loss made from 10~km fiber.

The QKD systems introduced here represents almost an order of magnitude increase in the secure key distribution rate, from previously 1.9~Mb/s \cite{dynes16} to 13.72~Mb/s. Furthermore this is complete system, with all layers of the system able to operate at this speed without bottlenecks. This scaled up key rate makes it viable to use the systems as a backbone in a large quantum network, allowing multiple users to share the link simultaneously while providing each with Mb/s key rates for the first time. This is especially relevant as most recent deployments of QKD systems are focusing on network configurations, due to their inherent advantages in redundancy and reach.

When used in a dedicated link configuration the system can provide key rates to allow for the most secure one time pad encryption of all typical types of communication, including voice, video, medical and financial data. This ability should allow for the ultra high security provided by QKD to be deployed in a wide range of applications.

\ifCLASSOPTIONcaptionsoff
  \newpage
\fi

\bibliographystyle{IEEEtran}
%\bibliography{./ref}


\end{document}